\documentclass[galaxies,article,accept,pdftex,moreauhtors]{Definitions/mdpi} 
\firstpage{1} 
\pubvolume{1}
\issuenum{1}
\articlenumber{0}
\pubyear{2025}
\copyrightyear{2025}
\externaleditor{Firstname Lastname} 
\datereceived{12 May 2025 } 
\daterevised{4 June 2025 } 
\dateaccepted{9 June 2025 } 
\datepublished{ } 
\hreflink{https://doi.org/} 

\newcommand*\aap{A\&A}

\newcommand*\araa{ARA\&A}

\Title{Post-AGB Binaries as Interacting Systems}

\TitleCitation{Post-AGB Binaries as Interacting Systems}

\Author{Hans Van Winckel 
 \orcidA{}}

\AuthorNames{Hans Van Winckel}
\AuthorCitation{Van Winckel, H.}

\address[1]{Institute of Astronomy, Department of Physics and Astronomy, KU Leuven, Celestijnenlaan 200D, \linebreak  3010 Leuven, Belgium; hans.vanwinckel@kuleuven.be}

\abstract{We present recent progress in our understanding of the physical interaction mechanisms at work in evolved binaries of low-to-intermediate initial mass, which are surrounded by a stable disc of gas and dust. These systems are known as post-asymptotic giant-branch (post-AGB) binaries, but recently, it has been shown that some systems are too low in luminosity and should be considered as post-red-giant branch (post-RGB) instead. While the systems are currently well within their Roche lobe, they still show signs of active ongoing interaction between the different building blocks. We end this contribution with some future research plans.  }


\begin{document}


\section{Introduction}
Post-asymptotic giant-branch (post-AGB) stars are descendants of low-to-intermediate mass main-sequence stars. This class is normally limited to objects that have already left the AGB but~are not yet hot enough to ionize their surroundings, e.g.,~\citep{vanwinckel03}. 

\begin{figure}[H]

\includegraphics[width=.5\textwidth]{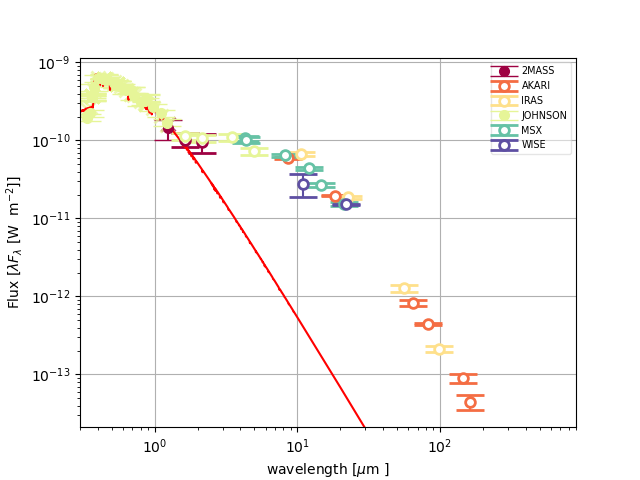}
\includegraphics[width=.5\textwidth]{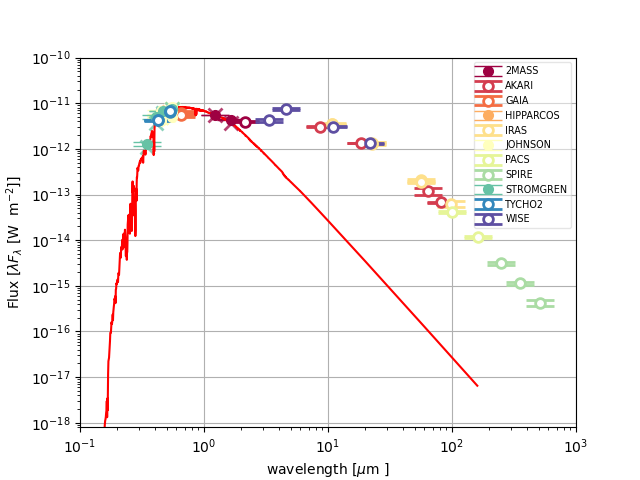}

\caption{The 
 spectral energy distribution of 3Pup (\textbf{left}), e.g.,~\citep{miroshnichenko20}, a~well studied FS CMa star and the post-AGB star HD131356 (\textbf{right}). As~both central stars do not have a current dusty mass outflow, the~near-IR excess and general SED is indicative of the presence of a stable dusty disc, e.g.,~\citep{kluska22}.\label{fig:sed}}
\end{figure}

Binaries among these post-AGB objects have a common observational property: They display a clear near-infrared excess, indicating that circumstellar dust must be close to the central star(s) (see Figure~\ref{fig:sed}). It is now well established that this indicates the presence of a stable circumbinary disc of gas and dust, e.g.,~\citep{vanwinckel18}. The~luminous evolved post-AGB primary has an unevolved stellar companion, which has a minor contribution to the energy \mbox{budget \citep{oomen18}}.
The SEDs are very similar to the SEDs of FS CMa objects (see Figure~\ref{fig:sed}), as these objects are also surrounded by dusty discs. During~the Almaty 2024 conference on hot stars with circumstellar material, we have seen several detailed studies of FS CMa objects (see other papers of this special volume).

\vspace{-9pt}

In this contribution, I give an overview of the recent progress in our research on post-AGB disc sources. For~FS CMa objects, as well as binary post-AGB stars, there is ample evidence that the creation and evolution of the discs and their impact on the orbital properties of the systems are key missing ingredients in our understanding of their~evolution.

The sample of post-AGB stars is more homogeneous than FS CMa stars or B[e] stars. In~our galaxy, a~sample of about 85 of these disc objects was identified \citep{kluska22}, which is about 25--30\% of all known galactic post-AGB stars! Also, in the Large and Small Magellanic Clouds, disc sources represent about a third of the known population of optically bright post-AGB stars \citep{kamath14a, kamath15a}.  Given the numbers, disc creation must be a mainstream process. The~creation and evolution of circumbinary discs are important ingredients in binary stellar evolution for a significant number of~objects. 

Some of these objects have luminosities lower than the tip of the Red-Giant Branch; they can be identified as dusty post-RGB stars \citep{kamath16} rather than post-AGB stars.
These luminosities can be determined from SED fitting combined with a
distance \mbox{measurement \citep{kluska22}}, although~they have large uncertainties due to substantial errors in both the distance measurement and the reddening determination. The best estimates are that 38/85 objects have a luminosity below the tip of the RGB \citep{moltzer25}. The~low luminosity for some variable objects could be confirmed via the period--luminosity--color relation of Population II \mbox{Cepheids \citep{manick19,mohorian24}}.  As these objects are structurally very similar, we keep the nomenclature of post-AGB binaries for all dusty evolved binaries in the context of this~paper.

In this overview paper, I give a status report on recent research on post-AGB binaries with their circumbinary dusty discs. This~paper can be seen as an update of the chapter by~\cite{vanwinckel18}. This~paper starts with a review on the orbital properties, after~which we focus on dissecting the common building blocks. We also review the research on  ongoing interaction processes as it is now clear that these objects must be seen as interacting binaries despite the fact that the luminous object is well within the Roche Lobe. We end with possible future avenues of research on these interesting~systems.

\section{Orbital Distribution of Post-AGB~Stars}

While the first single-lined spectroscopic binaries (SB1) among post-AGB disc sources were serendipitously detected, we started a very systematic radial velocity monitoring program \citep{vanwinckel15} with our 1.2 m Mercator telescope using the HERMES spectrograph \citep{raskin11}. This led to the observational confirmation that the disc sources are binaries with periods in the 100--3000-day range with remarkable eccentric orbits. The~period distributions, as well as high eccentricities, are difficult to align with theoretical~expectations.

\begin{figure}[H]
\includegraphics[width=0.99\textwidth]{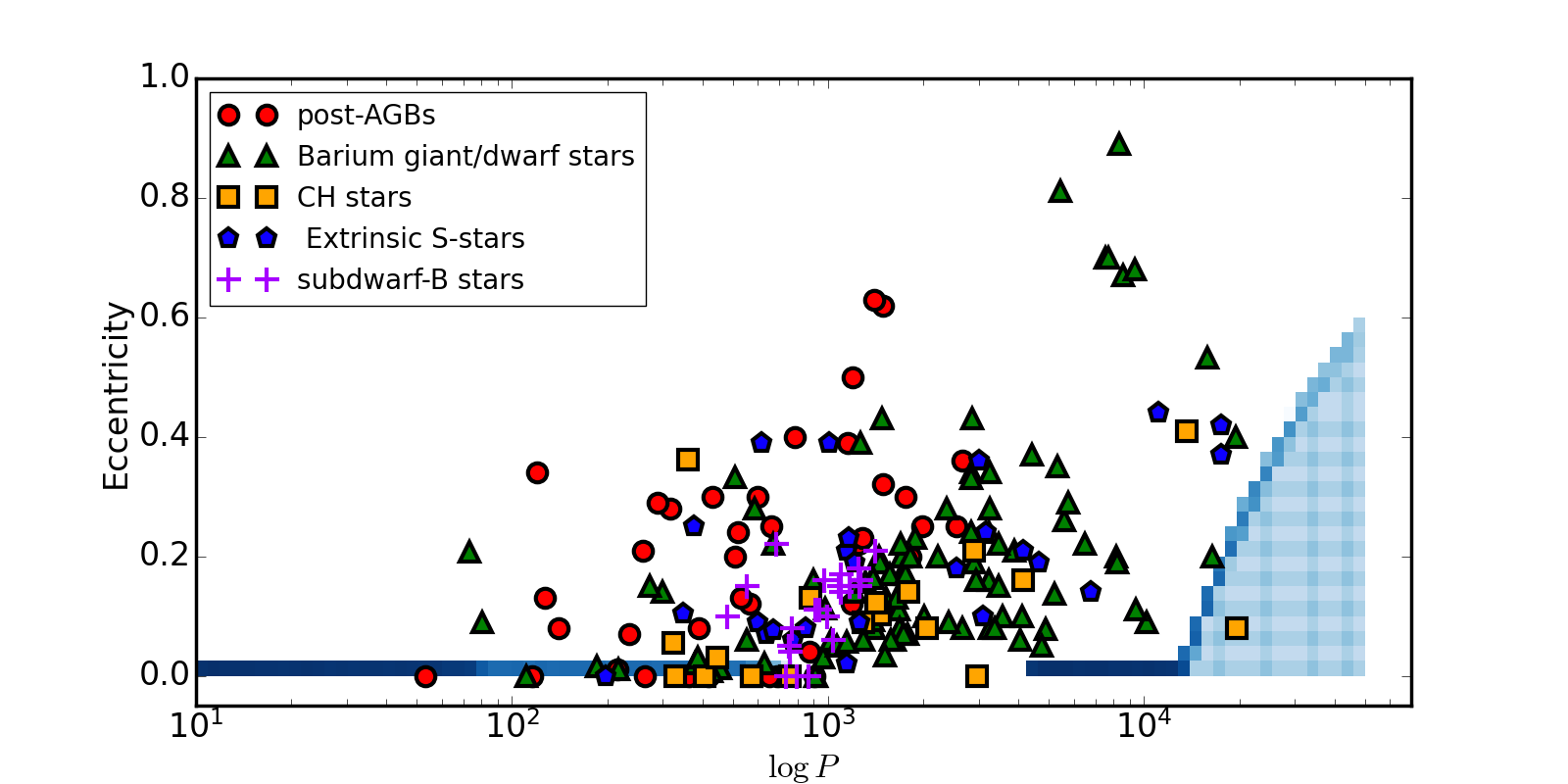}
\caption{The e–log(P) diagram of evolved binaries, including post-AGB stars \citep{oomen18}; systems with white dwarfs such as Ba, CH, and CEMP-s; and extrinsic S stars \citep{jorissen16, jorissen19, escorza19, hansen16} and~wide sdB stars \citep{vos17}. Population synthesis (SeBa;~\cite{toonen12} results are shown in blue, with~darker areas indicating the most probable~outcomes.\label{fig:bin}}
\end{figure}

In Figure~\ref{fig:bin}, we illustrate the eccentricity--period diagram of spectroscopically observed evolved binaries, where the most evolved stellar component is either a cold white dwarf, a~sub-dwarf B star, or a post-asymptotic giant-branch (post-AGB) star. Those color-coded in blue are also theoretical predictions, and the darker the blue, the~more likely the predicted position. The~majority of observed orbits fall within the period range least predicted by theory! Figure~\ref{fig:bin} illustrates a well-documented astrophysical challenge: The theoretical predictions are in stark contrast with the observed orbital periods and eccentricities, revealing a strong tension between stellar interaction theory and observations, e.g.,~\citep{pols03, bonacic08, abate15, nie17, oomen18}. Note that the very long periods are hard to detect spectrocopically, and the observational sample is biased against these long periods. This tension is also present in the recently quantified astrometric orbits of white dwarf binaries, as detected using GAIA DR3 \citep{yamaguchi25}.

Clearly, fundamental ingredients are missing in current theories, and~one of the ultimate goals in our research community is to combine a wide range of state-of-the-art observational and computational methods to lift this tension. 
\section{Dissecting the Building~Blocks}

As the galactic post-AGB binaries are located at typical distances of one to several kiloparsecs, the~angular sizes of the circumbinary material is expected to be small, and~state-of-the-art high-spatial-resolution techniques are needed to resolve them and unveil both the compact nature of the inner region and the strong similarity between the circumbinary discs and protoplanetary discs (PPDs) around young stellar objects (YSOs), e.g.,~\citep{kluska19, hillen17, corporaal21, corporaal23, andrych23, andrych24}. 

Based on their IR color--color diagram and spectral energy distribution, discs are the classified into categories \citep{kluska22}: 
Full discs are dust discs with an inner radius at the sublimation radius extending to radii of ~500AU. In~about 10\% of the discs, however, the~dust-free cavity is larger than the dust sublimation radius, resembling the transition discs seen in young stellar objects \citep{corporaal23}. Another SED category entails objects that have a very high infrared luminosity, much larger than what is obtained by redistributing the energy lost by circumstellar reddening into the infrared region. This is interpreted as a result of an aspect angle, which is close to edge-on sources. For~these sources, the~optical flux is dominated by scattered light rather than reddened direct light. Finally, binaries with very little IR excess were also identified (see \citep{kluska22} for a full description).

CO interferometric observations and the modeling of a sample of seven sources show that gas discs extend further and up to
1000 AU in size \citep{gallardocava21}. In~addition, slow outflows are observed, expanding at
5--10 km s$^{-1}$. Two different type of systems were identified:  disc-dominated objects, where more than 85\% of the total nebular mass is contained within the disc \citep{bujarrabal16, bujarrabal17, bujarrabal18, gallardocava21}, and~outflow-dominated subclass
features with rotating discs that are less massive. This segregation is also evident in their single-dish CO spectra, where disc-dominated sources exhibit narrow lines with faint wings, while outflow-dominated sources display narrow lines accompanied by broad complex wings \citep{bujarrabal13b}.

\begin{figure}[H]
\includegraphics[width=\textwidth]{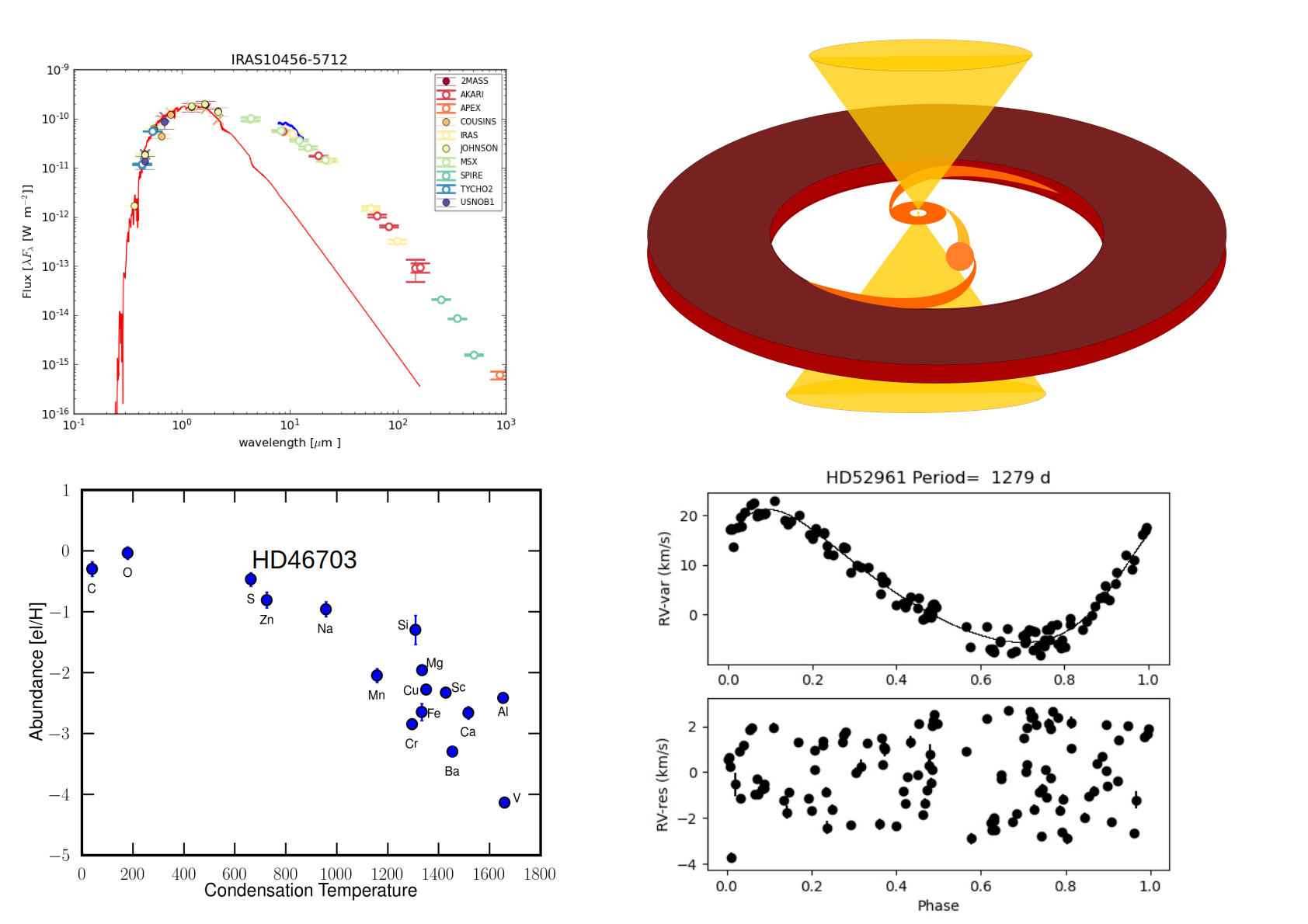}
\caption{\textbf{Top left}: The spectral energy distribution of a post-AGB star showing a typical infrared excess of a disc \citep{kluska22}. \textbf{Bottom left}: The strong photospheric chemical anomaly: elements with a high dust condensation temperature are strongly depleted, e.g.,~\citep{vanwinckel18}. \textbf{Top right}: A diagram with the most important building blocks of these objects and a circumbinary disc of gas and dust, with an~accretion flow towards both components and a jet-like outflow created around the companion \citep{bollen22}. For~clarity, the large circumbinary disk wind is not shown. \textbf{Bottom right}: A typical radial velocity curve fitted with a Keplerian orbit. The~residuals are mainly determined by stellar pulsations \citep{oomen18}. 
\label{fig:components}}
\end{figure}

In Figure~\ref{fig:components}'s top-right panel, we illustrate the building blocks that contain a central binary with an evolved primary and a dim unevolved main-sequence companion while both are surrounded by a stable circumbinary disc of gas and dust \citep{vanwinckel18}.

\vspace{-6pt}

In what follows, we focus on the strong interactions between the different components that are traced, on the one hand, by the detection of jets (see Section~\ref{section:jets})  and, on the other hand, by the detection of chemical anomalies in the photospheric surface abundances of the evolved primary (see Section~\ref{section:depletion}).  Although~the building blocks may be identified, it is unclear how the physical interactions between the blocks work and~evolve.

\section{Ongoing Interaction~Processes}
\unskip

\subsection{Jets as MHD Disc~Winds} \label{section:jets}
\textls[-15]{The detection of disc winds is founded on a decade-long time series of high-resolution spectra we have obtained and are obtaining with the 1.2 m Flemish Mercator telescope \citep{raskin11,vanwinckel10}. }

\begin{figure}[H]

\includegraphics[width=.45\textwidth]{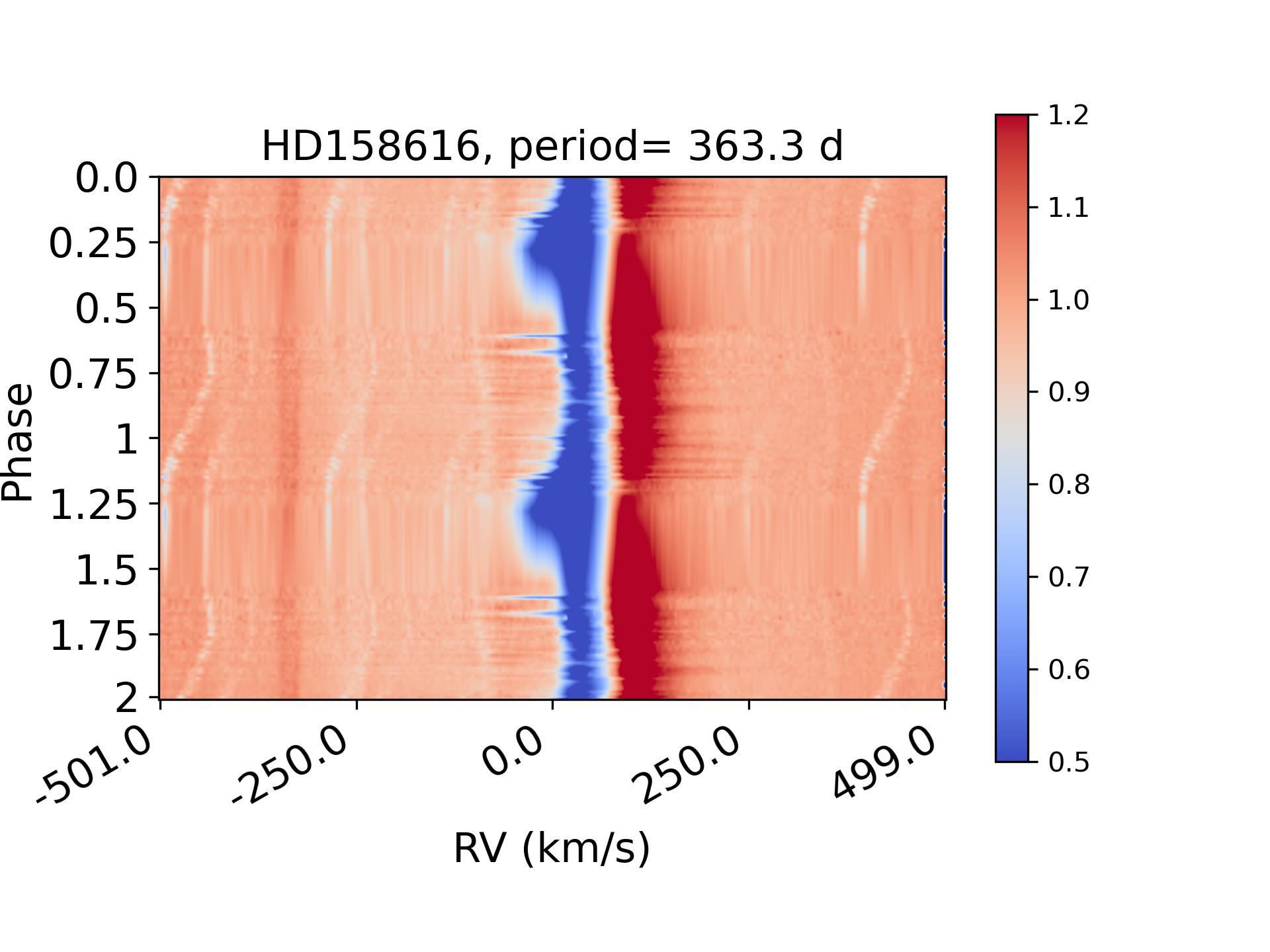}~~~~~
\hspace{-10pt}\includegraphics[width=.45\textwidth]{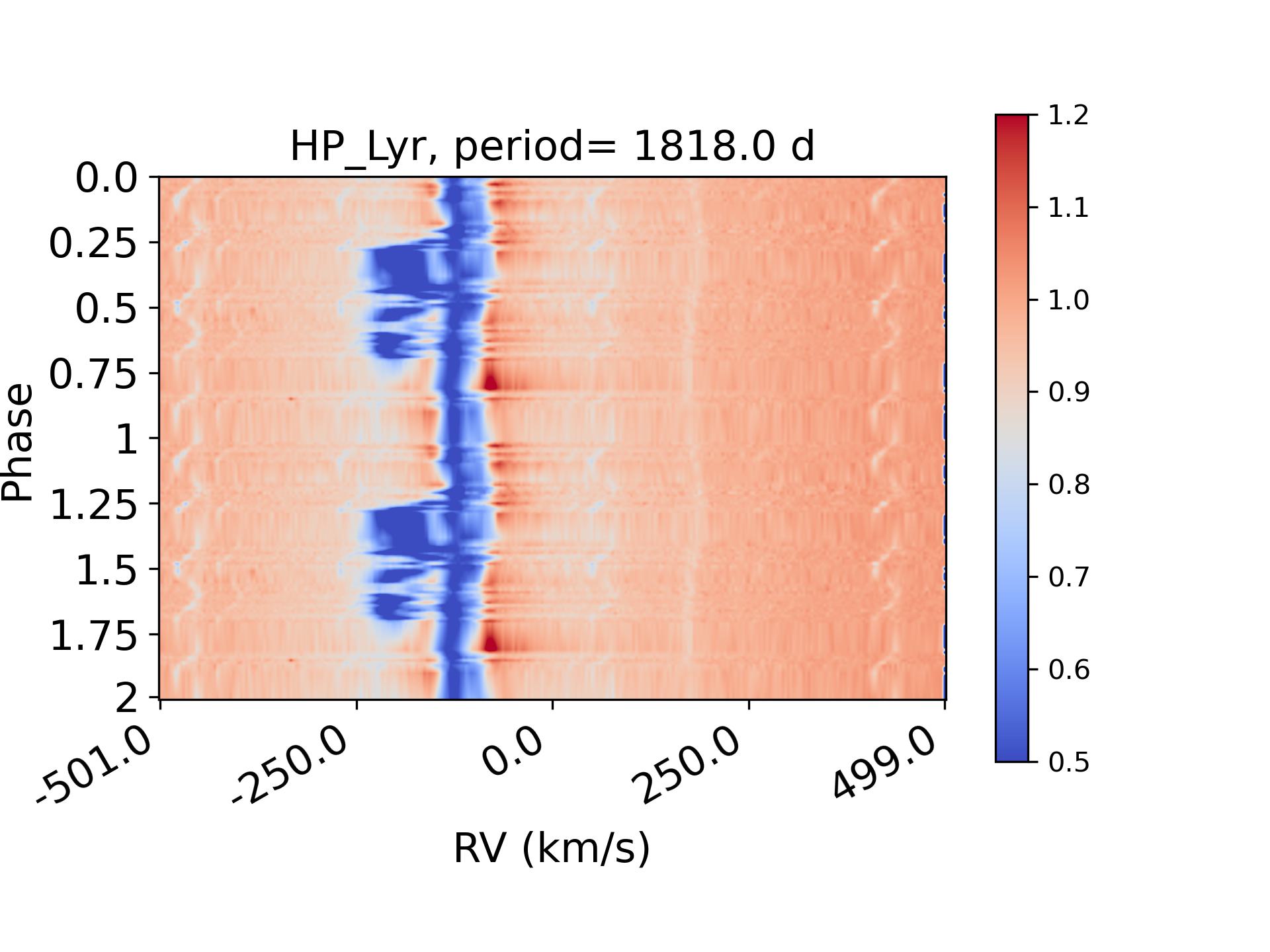}

\includegraphics[width=.45\textwidth]{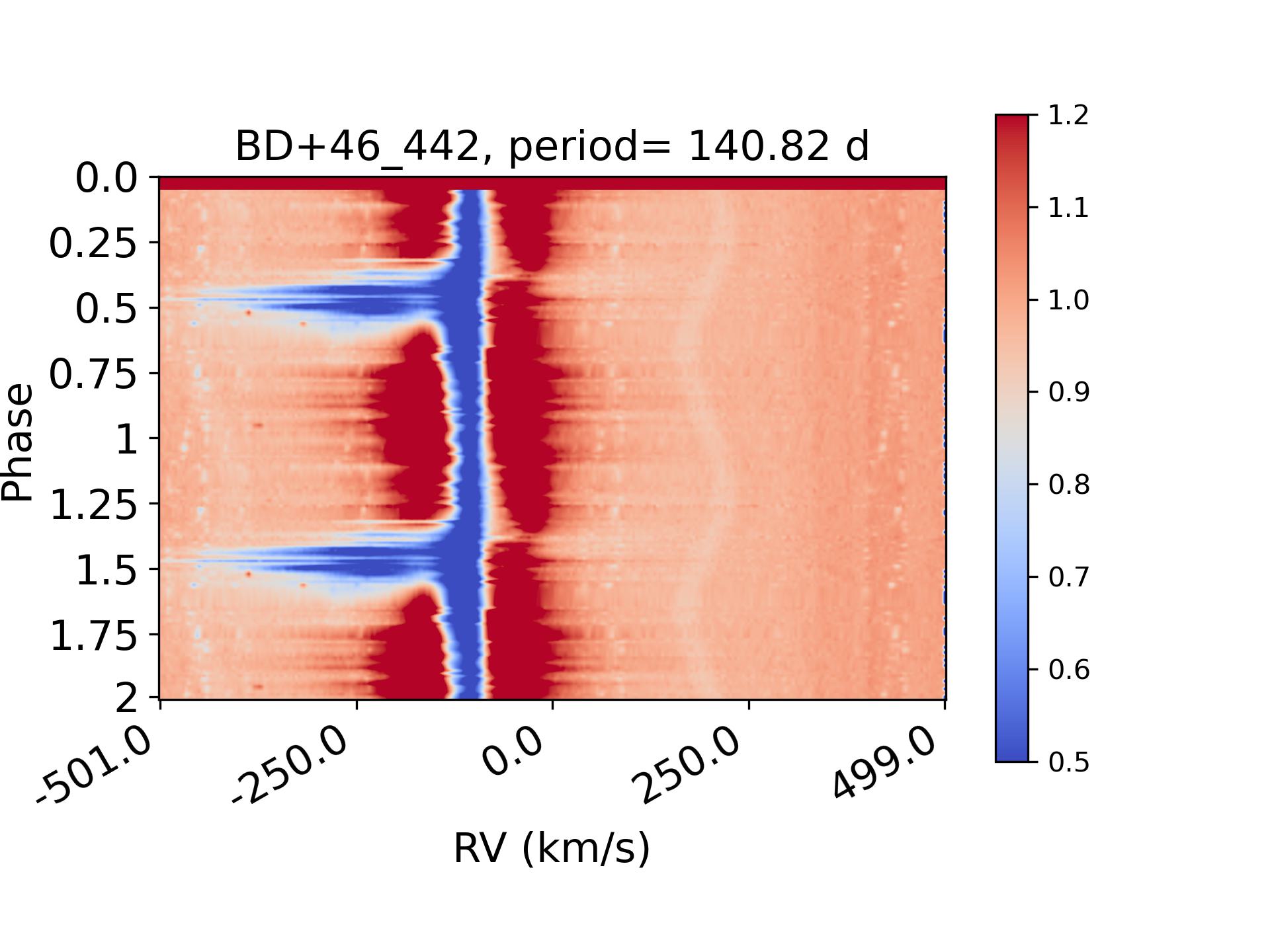}
\includegraphics[width=.45\textwidth]{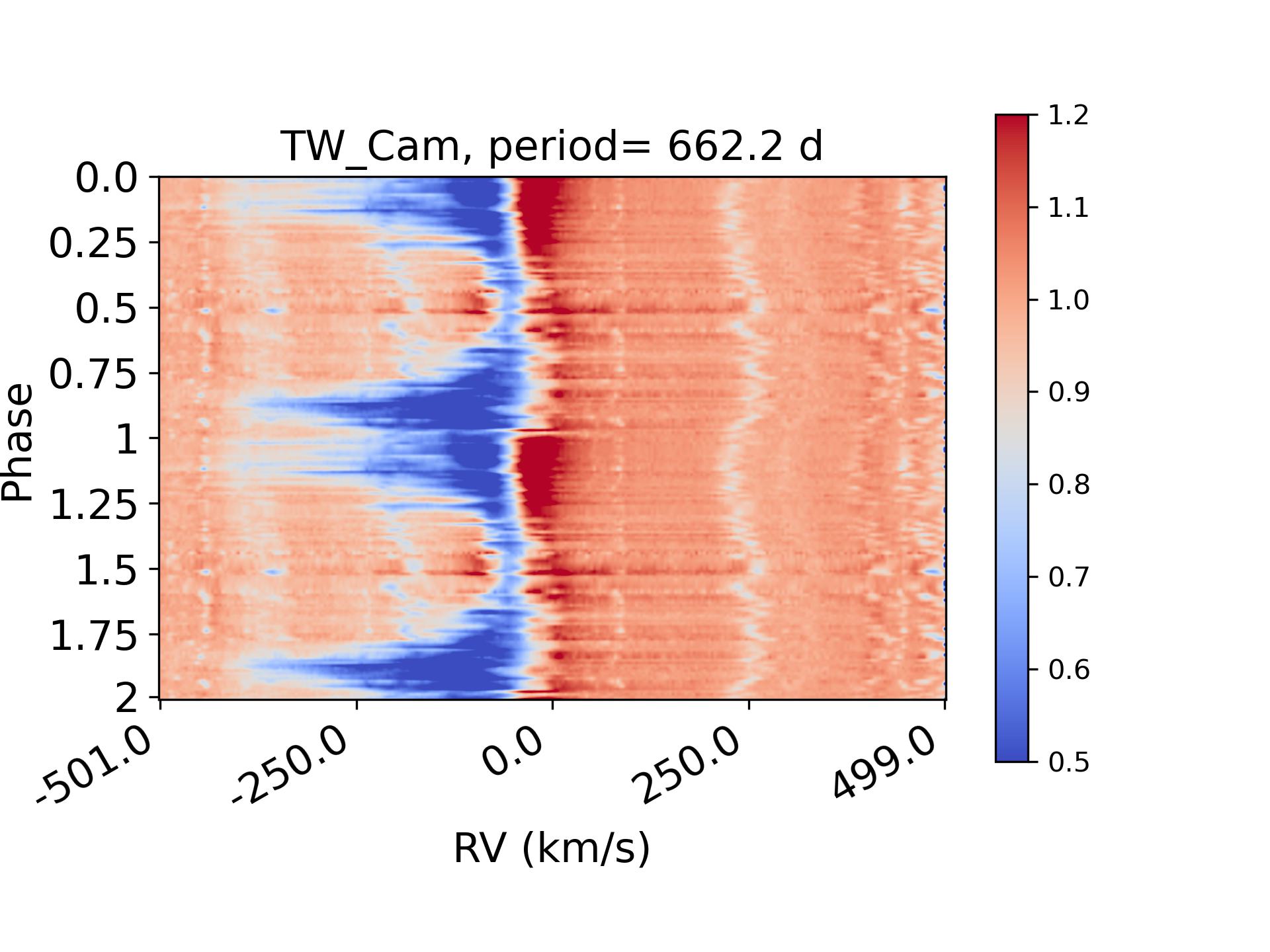}

\caption{Dynamic spectra around the $H_{\alpha}$ region. The~spectra are phase-folded on the binary orbital period, and for clarity, we show two cycles. The zero phase is arbitrary. We see a clear high-velocity absorption component when the companion appears in front of the luminous primary, e.g., (\citep{deprins24}, and references therein).\label{fig:halpha}}
\end{figure}

When we phase-fold the spectra (see Figure~\ref{fig:halpha}) on the orbital period and~focus on the $H_{\alpha}$ profile, we detect an absorption component that appears around a superior conjunction, when the primary star is at its furthest point from us and its companion is in front \citep{gorlova15, bollen17, bollen19, bollen20, bollen21, bollen22, verhamme24,deprins24}. It is now well established that this is due to a high-velocity biconical outflow or jet, which is launched from the accretion disc around the secondary point. When the companion moves in front of the primary, one of the bicones gradually enters the line of sight towards the primary. This creates an absorption component when continuum photons from the background post-AGB primary are selectively removed from the line of sight. The~orbital properties, the~aspect angle, and the jet geometry determine which part of the jet is occulting the line of sight towards the primary at any given time. These systems therefore provide unique opportunities in which time-resolved spectra during orbital motion scan the biconical flow very close to the launching engine within~or just outside the accelerating zone. These data can be used to investigate the dynamics of the jet and the structure of the jet. Thanks to our systematic 15-year-long monitoring, we have data for about 35 objects that are well sampled in the orbital phase, despite the fact that the orbits are long (\textasciitilde100 to 2500 days).\vspace{-6pt}

We developed an analysis approach in several steps, with~the aim to produce physical self-consistent models that mimic phase-dependent line profiles. First, we focused  on developing a purely geometric spatio-kinematic model to fit the observed time series. This was followed by a line-tracing radiative transfer module to~quantify the total mass loss in the~jets. 

The common results of the 16 objects analyzed with geometric models to date are as follows: The jets are found to be wide (>30 degrees) and display an angle-dependent density structure with a dense and slower outer region near the jet cone, and~they have a fast inner part along the jets' symmetry axes. The~de-projected outflow velocities can be evaluated as the escape velocities, and they confirm that the companions are main-sequence stars. Many objects show a high-mass-jet-ejection rate ($10^{-4}$--$10^{-7}\,M_{\odot}\,yr^{-1}$), which should always be less than the accretion rate relative to the accretion disc with a typical value of 10\% ejection efficiency. These accretion rates are much higher than the typical mass loss rates of post-AGB stars. The~circumbinary disc is therefore likely the feeding reservoir of the circum-companion accretion~disc. 

 Next, we moved away from geometric modeling and included self-similar physical jet models to~characterize the properties of these outflows and the physical conditions and jet-launching mechanisms inside their accretion discs. These models use super-Alfv\'enic self-similar MHD solutions \citep{jacquemin-ide19}. Using a self-similar ansatz, they consistently solve the equations of resistive single-fluid 2D MHD for the combined disc--jet system, e.g.,~\citep{casse00}. This gives the relative scaling of the wind geometry, density, and kinematics.  The~three important parameters that characterize these models are the accretion disc ejection efficiency; magnetization; and~the scale height of the accretion disc, which launches the jet, e.g.,~\citep{verhamme24, deprins24}. We adapted our framework such that the MHD disc models are included in a binary setting, with the primary being the post-AGB star, which serves as a background light source. This initial parameter study shows that these MHD models indeed produce similar synthetic dynamic spectra. In~\citep{deprins24}, we expand our framework even further and use five different MHD solutions as inputs to synthesize the spectral time series of the $H_{\alpha}$ line. Relative to jets launched from young stellar objects (YSOs), all five evolved targets prefer winds with higher ejection efficiencies, lower magnetization, and thicker~discs. 
 
 We concluded that current cold MHD disc wind solutions can explain many of the jet-related $H_{\alpha}$ features seen in post-AGB binaries, though~systematic discrepancies remain. This includes, but~is not limited to, overestimated predicted rotation. Jet rotation csould also induce a red-shifted component onto the line profile, which is never detected \citep{deprins24}.  The~consideration of thicker discs and the inclusion of irradiation from the post-AGB primary, leading to warm magneto--thermal wind launching, might alleviate these phenomena. Additional challenges are the needed high accretion rates that would imply very-short post-AGB circumbinary disc lifetimes, which is in contradiction with the fact that we detect so many of them! 

\subsection{Chemical~Depletion}
\label{section:depletion}
Another outstanding feature is that many objects show a strong chemical anomaly in their photosphere: Elements that easily stick to dust grains (refractory elements having high dust condensation temperatures, like Fe, Ti, or Sc) are strongly depleted, while volatile elements (like Zn and S) are not affected (see Figure~\ref{fig:components}; bottom right). The~depletion \mbox{patterns \citep{maas05, giridhar05, rao14, kamath19, oomen19, mohorian25}} can be explained as a result of the selective accretion of only the volatiles onto the surface of the post-AGB star, while the refractory elements remain trapped in the circumbinary disc \citep{waters92}. This chemical composition resembles the chemical composition of gas in the interstellar medium. The condensation temperatures in Figure~\ref{fig:components} are determined by  chemical equilibrium computations from a solar mixture and at a constant pressure, meaning a oxygen-rich environment, e.g.,~\citep{lodders03}. In order to change the chemical composition of the photosphere of the evolved component, an~interaction process is required, which separates the gas from the dust in the circumbinary disc, followed by the selective accretion of this clean gas only.
The photospheric chemical composition is a mixture of the original composition, altered by dredge-up processes during evolution and the clean accreted gas. Ref.~\cite{kluska22} showed that objects with a transition disc are more depleted than the bulk of the full disc sources.
This was confirmed by~\cite{mohorian25}, which also showed that saturated abundance profiles~prevail.

The strong link between depletion and infrared excess colors and, thus, disc morphology points toward a mechanism that
separates the gas from dust within the disc while shaping the
disc's~structure. 

Similar gas--dust separation and accretion processes are observed in young stellar objects (YSO), again highlighting the strong similarity between the discs around evolved binaries and the planet forming discs around young stars. The~discs around evolved stars can therefore be seen as protoplanetary discs of the second generation, e.g.,~\citep{kluska22}. 
A natural explanation for the gas--dust separation would be the presence of a third component in the system that carves the disc hole in the dust by acting
as a filter, trapping the dust in the outer disc but still allowing the
depleted gas to be accreted onto central stars. This is also an advocated process in YSOs.  It was proposed by \citep{kama15} that giant planets (with masses of
1–10 M$_{jup}$) could be responsible for the observed depletion pattern
in Herbig~stars.

We also conclude that the chemical photospheric depletion process is intimately connected to the interaction between the circumbinary disc and the inner~binary.

\section{Conclusions}

In this contribution and in line with the Almathy 2024 conference on “Hot Stars. Life with Circumstellar Matter”, I focused on the recent progress in our understanding of interaction processes between all components within binary post-AGB objects. While the evolved component is typically well within its Roche lobe, the interactions between the different components are strong. We have shown that these common components comprise a central binary with an evolved component and a main-sequence secondary, a~large circumbinary disc, and a biconical outflow powered by an accretion disc around the~companion.

It is now well established that the creation and evolution of the circumbinary discs around post-AGB binaries play a lead role in the final evolution of binaries with low-to intermediate initial mass.  The~circumbinary discs are transient phenomena, but~their impact on the central stars needs to be fully understood in order ~to investigate if they can help solve the puzzle depicted in Figure~\ref{fig:bin}. 

Very similar discs are found around B[e] stars or FS CMa stars, e.g.,~\citep{kluska20b}, and  future research should also focus on the similarities of the physical processes of these discs in very different energetic environments and evolutionary phases: from~the planet forming discs around young systems to the second generation of proto-planetary discs around evolved systems all over the HR diagram. High-spatial-resolution data will be crucial, as for the time being, this type of research is mainly observationally~driven. 

\vspace{6pt} 

\funding{FWO is acknowledged for the IRI funding with contract number I000325N.} 

\dataavailability{The archive of the HERMES spectrograph on the Mercator observatory will become available on http://www.mercator.iac.es/ in the course of 2025-2026} 

\acknowledgments{
This study is based on observations made with the Mercator Telescope, operated on the island of La Palma by the Flemish Community, at~the Spanish Observatorio del Roque de los Muchachos of the Instituto de Astrofísica de Canarias. This study is also based on observations obtained with the HERMES spectrograph, which is supported by the Research Foundation---Flanders (FWO), Belgium, the~Research Council of KU Leuven, Belgium; the~Fonds National de la Recherche Scientifique (F.R.S.---FNRS), Belgium; the~Royal Observatory of Belgium; the~Observatoire de Gen\`eve, Switzerland; and the Th\"uringer Landessternwarte Tautenburg, Germany.  
 is a pleasure to acknowledge the long-time collaborator Devika Kamath and Jacques Kluska, as well as the very pleasant and fruitful collaboration with Valentin Bujarrabal, Javier Alcolea, Iv\'an Gallardo Cava, Onno Pols, Jonathan Ferreira, Michel Hillen, Nadya Gorlova, Ana Escorza, Shreeya Shetye, and Orsola De Marco in this research context. Moreover,  the specific research on binary post-AGB stars would not have been possible without the talented PhD students we were lucky enough to host at the KULeuven, Nijmegen, and Macquarie universities and who mainly worked on this research topic: Pieter Deroo, Thomas Maas, Joris Vos, Rajeev Manick, Glenn-Michael Oomen,  Dylan Bollen, Olivier Verhamme, Kateryna Andrych, Maksym Mohorian, Megna Mukesh Menon, Casper Moltzer, and Toon De~Prins.
}
\conflictsofinterest{The author declares no conflicts of interest.} 

\reftitle{References}
\begin{adjustwidth}{-\extralength}{0cm}





\PublishersNote{}
\end{adjustwidth}
\end{document}